\documentclass[aps,twocolumn,showpacs,superscriptaddress,amsmath,amssymb]{revtex4}

\usepackage{graphicx}
\usepackage{dcolumn}
\usepackage{bm}

\begin{document}

\title{Bandgap and Band Offsets Determination of Semiconductor Heterostructures using Three-terminal Ballistic Carrier Spectroscopy\\}

\author{Wei Yi}
\email{weiyi@seas.harvard.edu}
\affiliation{School of Engineering and Applied Sciences, Harvard University, Cambridge,
Massachusetts 02138
}

\author{Hong Lu}
\affiliation{Materials Department, University of California, Santa Barbara, California 93106
}

\author{Yong Huang}
\affiliation{Center for Compound Semiconductors and School of Electrical and Computer Engineering, Georgia Institute of Technology, Atlanta, GA 30332
}

\author{Michael A. Scarpulla}
\affiliation{Materials Department, University of California, Santa Barbara, California 93106
}

\author{Jae-Hyun Ryou}
\affiliation{Center for Compound Semiconductors and School of Electrical and Computer Engineering, Georgia Institute of Technology, Atlanta, GA 30332
}

\author{Arthur C. Gossard}
\affiliation{Materials Department, University of California, Santa Barbara, California 93106
}

\author{Russell D. Dupuis}
\affiliation{Center for Compound Semiconductors and School of Electrical and Computer Engineering, Georgia Institute of Technology, Atlanta, GA 30332
}

\author{Venkatesh Narayanamurti}
\affiliation{School of Engineering and Applied Sciences, Harvard University, Cambridge,
Massachusetts 02138
}

\date{\today}

\begin{abstract}
Utilizing three-terminal tunnel emission of ballistic electrons and holes, we have developed a method to self-consistently measure the bandgap of semiconductors and band discontinuities at semiconductor heterojunctions without any prerequisite material parameter. Measurements are performed on lattice-matched GaAs/Al$_{x}$Ga$_{1-x}$As and GaAs/(Al$_{x}$Ga$_{1-x}$)$_{0.51}$In$_{0.49}$P single-barrier heterostructures. The bandgaps of AlGaAs and AlGaInP are measured with a resolution of several meV at 4.2~K. For the GaAs/AlGaAs interface, the measured $\Gamma$ band offset ratio is 60.4:39.6 ($\pm2$\%). For the GaAs/AlGaInP interface, this ratio varies with the Al mole fraction and is distributed more in the valence band. A non-monotonic Al composition dependence of the conduction band offset at the GaAs/AlGaInP interface is observed in the indirect-gap regime.
\end{abstract}


\maketitle

Among the most important properties of semiconductor materials are their energy gaps and the relative alignment of the energy band edges at the heterojunction (HJ) interface between two dissimilar semiconductors, i.e. the way in which the total bandgap difference distributes between the conduction band discontinuity $\Delta E_{C}$ and the valence band discontinuity $\Delta E_{V}$. An accurate knowledge of such properties is crucial for the design of heterostructure devices widely used in high-speed and power electronics, photonics, and energy conversion. 

Numerous efforts have been devoted to measure the bandgap and band offsets. Bandgaps are measured mostly with optical spectroscopies such as absorption, photoluminescence (PL), photoluminescence excitation (PLE), and ellipsometry~\cite{yu-cardona:book}. Band offset measurements can be divided into three categories: electrical techniques such as thermionic emission and capacitance-voltage (C-V); optical techniques such as absorption, PL, and PLE; and photoelectron techniques such as x-ray photoelectron spectroscopy (XPS) and ultraviolet photoelectron spectroscopy (UPS)~\cite{yu-mccaldin-mcgill:book}. However, each of these methods has certain limitations or weakness. Absorption requires values of effective masses and parabolic band structure to measure the band offset~\cite{dingle:prl:33:827}. PL and PLE are often subject to interference from competing optical transitions due to strain splitting, phonon replicas, and impurities~\cite{ritter:pssb:211:869}. Thermionic emission requires current-voltage (I-V) measurements at different temperatures to extract the activation energies over barriers, and it does not work at low temperatures~\cite{batey:jap:57:484}. In C-V measurements, detailed device simulations are needed to consider the effect of deep levels in the barrier layer~\cite{thooft:apl:48:1525}. XPS and UPS are limited to measuring the valence band offests~\cite{yu-mccaldin-mcgill:book}. With a three-terminal device configuration, Ballistic Electron Emission Microscopy (BEEM) and Spectroscopy (BEES) utilize ballistic injection of electrons/holes to probe the band offsets of HJs buried beneath a metal-semiconductor (m-s) interface~\cite{beemreviews}. To measure the genuine barrier heights, a delta-doping is needed to reach a flat-band condition in the heterostructure, which adds another uncertainty due to the error in the doping level~\cite{oshea:prb:56:2026}. Most of the aforementioned methods require separately designed \emph{n}-type and \emph{p}-type HJs to measure $\Delta E_{C}$ and $\Delta E_{V}$ independently, therefore the results are not necessarily self-consistent.

In this letter, we report that the bandgap of a semiconductor, as well as both $\Delta E_{C}$ and $\Delta E_{V}$ at a semiconductor HJ, can be measured on the same device. Previous BEEM work has unexceptionally relied on unipolar carrier transport, i.e. either electrons or holes are injected into an \emph{n}-type or \emph{p}-type Schottky contact respectively. In contrast, our method utilizes both electron and hole injection into the same Schottky contact~\cite{yi:prb:75:115333}. This enables measurement of the energy maxima in both the conduction band (CB) and the valence band (VB) of the semiconductor collector. A sum of these two values gives the bandgap of the corresponding consituent layer. Since the metal Fermi level at the m-s interface is used as the potential reference, the measured bandgap value is independent of the Fermi level pinning position. Moreover, such a method does not involve the generation and recombination of electron-hole pairs as in optical spectroscopies, so that it is free of exciton effects. In the previous report~\cite{yi:prb:75:115333}, however, only one material (Al$_{0.4}$Ga$_{0.6}$As) was studied and no band offest information was obtained. Here we demonstrate that this method allows an accurate measurement of the band offsets in the CB and the VB of a buried HJ over a wide range of constituent compositions without any knowledge of the material parameters.


The design of the semiconductor collector is similar to to a typical BEEM device, with an unintentionally-doped double-HJ single-barrier (SB) heterostructure epitaxially grown on a heavily doped (\emph{p}-type in this case) substrate~\cite{yi:prb:75:115333}. Such a design ensures a \emph{linear} band profile in the depletion region and hence a constant electric field (\emph{E}-field) which can be tuned by a collector bias $V_{C}$ (see Fig.~1). Importantly, a \emph{nonequilibrium} flat-band condition can always be reached accommodating precise measurement of barrier heights, eliminating the need of a delta-doping which may not be optimized~\cite{oshea:prb:56:2026}. The barrier layer effectively suppresses the majority carrier (hole) drift-diffusion current when a forward bias is applied to the Schottky diode. This makes it possible to measure the tunnel injected ballistic electron current. Cooling the device to low temperatures further suppresses the majority current due to thermionic emission or thermally assisted tunneling across the valence-band barrier.

Ternary AlGaAs and quaternary AlGaInP alloys are chosen as the model III-V compounds in our study for their mature growth techniques and technological importance. Lattice-matched GaAs/AlGaAs and GaAs/AlGaInP SB heterostructures are grown on Zn-doped \emph{p}-GaAs (100) substrates by molecular-beam epitaxy and metalorganic chemical vapor deposition, respectively. The growth sequence starts with a 500~nm \emph{p}-doped ($5\times10^{18}$~cm$^{-3}$) GaAs buffer layer (layer ``4'' in Fig.~1, and so forth). After growing a 45~nm undoped GaAs layer (``3''), the 50~nm barrier layer (``2''), either AlGaAs or AlGaInP, is grown. The growth is finalized with a 5~nm GaAs cap layer (``1'') for surface passivation. For a systematic study, samples with a wide range of Al mole fraction \emph{x} = 0.0, 0.1, 0.2, 0.3, 0.42, 0.6, 0.8 and 1.0 for Al$_{x}$Ga$_{1-x}$As; and \emph{x} = 0.0, 0.2, 0.35, 0.5, 0.7, 0.85, and 1.0 for (Al$_{x}$Ga$_{1-x}$)$_{0.51}$In$_{0.49}$P; are grown under exactly the same condition for each alloy system. Monolithic metal-base transistors (MBT) are fabricated, using planar Al/AlO$_{x}$/Al tunnel junctions as the tunnel emitters for ballistic carriers.  The Al/AlO$_{x}$/Al tunnel junctions are fabricated using a shadow-mask technique with details described elsewhere~\cite{wei-unpublished}. All the devices were characterized at 4.2 K with specimens immersed in liquid helium to suppress the thermionic current. Care was taken to eliminate the effect of the finite base contact resistance on the measured apparent barrier heights~\cite{wei-unpublished}.

\begin{figure}
\includegraphics[width=0.5\textwidth]{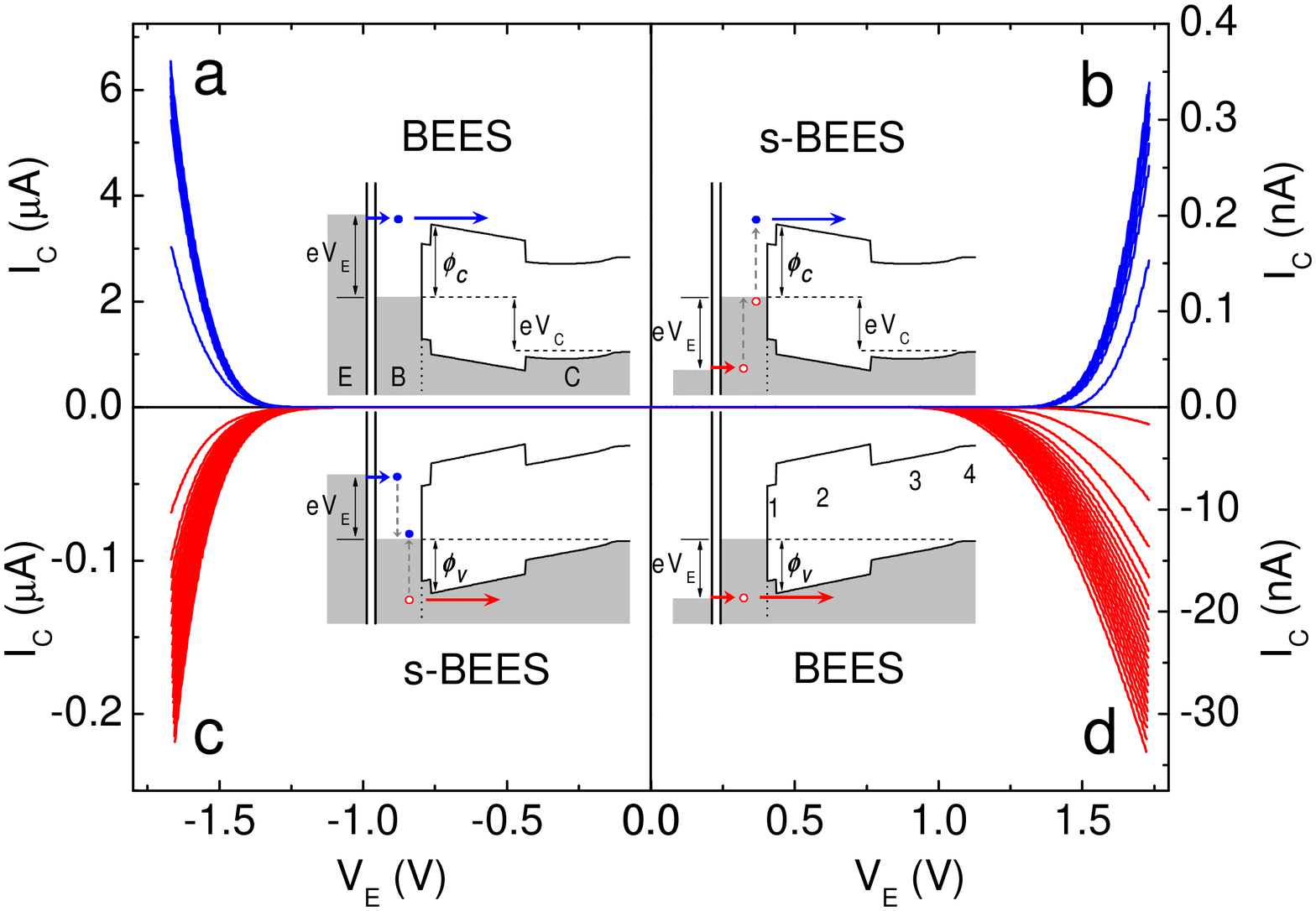}
\caption{Energy band diagrams (insets) and collector-current ($I_{C}$) spectra taken at 4.2~K illustrating the direct and secondary BEES processes. The \emph{p}-type heterostructure collector is composed of layer (1)-(4) as described in the text. Letter ``E'', ``B'', and ``C'' stand for emitter, base, and collector respectively. Emitter current $I_{E}\approx0.8(0.4)$~mA at emitter bias $V_{E}=\mp1.7$~V for this particular device.}
\label{fig1:rawdata}
\end{figure}


Ambipolar carrier injection is made possible using both the BEES and the secondary BEES (s-BEES) processes, as shown in Fig.~1. In an s-BEES process (also called ``reverse'' mode), e.g. hot electrons tunnel into the base from a negatively-biased emitter with the $p$-type collector unbiased or in reverse bias (Fig.~1(c)). The repulsive \emph{E}-field in the CB prevents them from being collected across the m-s interface. Rather, they lose their kinetic energy via electron-electron scatterings in the metal base and excite electron-hole pairs in an Auger-like process. As a result, hot holes are produced and some of them may be ballistically injected into the VB of the collector. These hot holes can be used to probe the valence-band barrier height ($\phi_{V}$). Under a forward $V_{C}$, the \emph{E}-field in the CB becomes attractive to collect hot electrons injected by the direct BEES process, and the conduction-band barrier height ($\phi_{C}$) is probed (Fig.~1(a)). Similar mechanisms apply for positive emitter biases (Fig.~1(b),(d)). The bandgap value of the barrier material is therefore determined by summing these two barriers. Despite the much smaller signal level, the energy resolution of s-BEES is superior to BEES owing to the much sharper current turn-on behavior of the two-step Auger process, i.e. in the near-threshold regime the collector transfer ratio $\alpha\equiv I_{C}/I_{E} \propto (eV_{E}-\phi_{V(C)})^{4}$~\cite{bell:prl:64:2679}, instead of a quadratic function $(eV_{E}-\phi_{V(C)})^{2}$ for BEES. As seen in Fig.~2(a) and (b), we found that the least squares linear fitting by an empirical function $\alpha^{1/4}=\alpha_{0}(eV_{E}-\phi_{V(C)})$, with only two free parameters, gives $\phi_{V(C)}$ with a typical energy resolution $<\pm2$~meV at 4.2 K ($k_{B}T\sim0.4$~meV), compared with $\sim$20 meV resolution in typical BEES fitted by a multi-valley Bell-Kaiser model. Note that the fitting process does not need band parameters, e.g. effective masses, making it applicable to materials for which such knowledge is not available. Note that the measured apparent barrier heights are subject to an image force lowering effect depending on the \emph{E}-field (i.e. $V_{C}$) in the undoped layers. Therefore, for each emitter bias polarity, a series of s-BEES spectra are measured under constant $V_{C}$ at 0.1~V intervals. The measured barrier heights  indeed change with $V_{C}$ (Fig.~2(c)). Maximum values of $\phi_{V}$ and $\phi_{C}$ are used to calculate the bandgap and band offsets, which are measured near the flat-band condition where the transition from hole to electron injection occurs.

\begin{figure}
\includegraphics[width=0.5\textwidth]{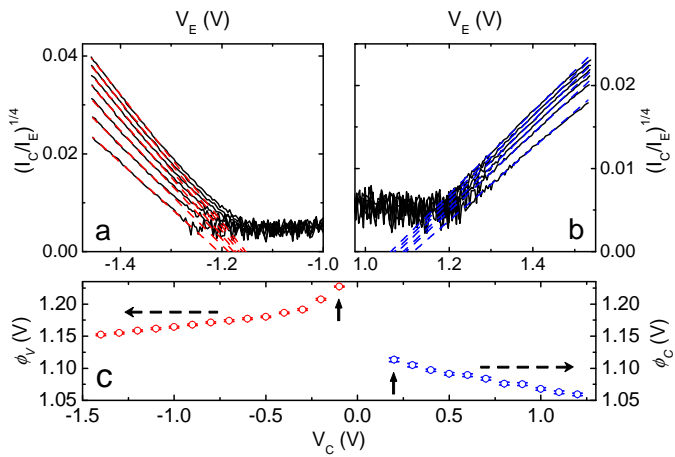}
\caption{Typical s-BEES taken at 4.2~K plotted as $(I_{C}/I_{E})^{1/4}$ under negative (a) and positive (b) emitter biases superimposed with linear fits (dashed lines). The spectra are spaced at 0.2 V $V_{C}$ intervals for clarity. (c) The measured barrier heights $\phi_{C}$ and $\phi_{V}$ as a function of $V_{C}$. Vertical arrows show the values used to determine the bandgap and band offsets.}
\label{fig2:datafit}
\end{figure}


The main results from the ternary AlGaAs alloy system are shown in Fig.~3. The measured $\phi_{V}$ and $\phi_{C}$ of Al-Al$_{x}$Ga$_{1-x}$As Schottky contacts, as functions of Al composition \emph{x}, are shown in Fig.~3(b) and (c), respectively. $\phi_{V}$ increases linearly in the full range of Al composition (0$<$$x$$<$1), which is expected because all the valence-band maxima of AlGaAs are located at the Brillion zone center. The VB offset $\Delta E_{V}(x)$ of GaAs/Al$_{x}$Ga$_{1-x}$As is obtained by $\Delta E_{V}(x)=\phi_{V}(x)-\phi_{V}(0)$, where $\phi_{V}(0)$ is the value for GaAs. Here it is assumed that the Fermi level pinning position at the m-s interface (wrt the vacuum level) remains unchanged for devices with different Al mole fraction. This assumption is validated by the fact that the measured $\phi_{V}$ shows a nearly perfect linear dependence on \emph{x}. A least squares linear fitting gives $\Delta E_{V}(x)=(0.57\pm0.01)x$~(eV). The CB offset $\Delta E_{C}(x)$ of GaAs/Al$_{x}$Ga$_{1-x}$As is derived from $\Delta E_{C}(x)=\phi_{C}(x)-E_{g}(0)+\phi_{V}(0)$, where $E_{g}(0)$ is the bandgap of GaAs (taken as 1.519 eV at 4.2K~\cite{vurgaftman:jap:89:5815}). The derived $\Delta E_{C}(x)$ increases linearly with \emph{x} in the direct regime, i.e. $\Delta E_{C}(x)=(0.87\pm0.01)x$~(eV) for $x$$<$0.42. A direct-indirect transition of the CB minima at $x=0.42$ is indicated by the adrupt slope change at this composition. In the indirect-gap regime, the CB offset $\Delta E_{C}(x)$ slowly increases with \emph{x}. The direct-gap $\Gamma$ band offset ratio $r\equiv\Delta E_{C}/\Delta E_{V}$ is found to be 60.4:39.6 ($\pm2$\%), which is very close to the consensus value of 60:40~\cite{missous:AlGaAsbook}. 

\begin{figure}
\includegraphics[width=0.4\textwidth]{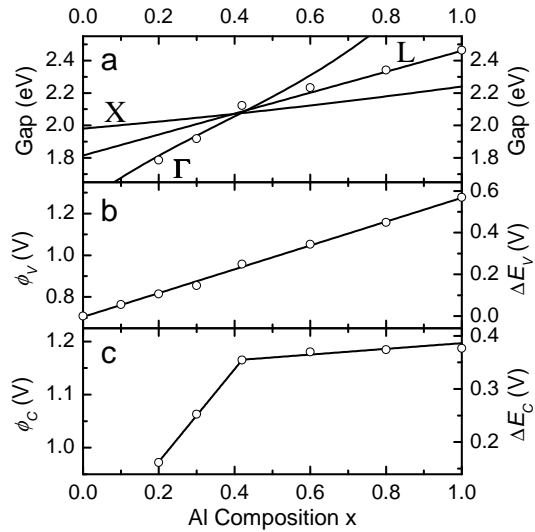}
\caption{Al mole fraction \emph{x} dependence (measured at 4.2 K) of (a) energy bandgap of Al$_{x}$Ga$_{1-x}$As (solid lines are from Ref.~\cite{vurgaftman:jap:89:5815}), (b) valence-band barrier $\phi_{V}$ and valence-band offset of GaAs/Al$_{x}$Ga$_{1-x}$As (solid line is a linear fit), and (c) conduction-band barrier $\phi_{C}$ and conduction-band offset of GaAs/Al$_{x}$Ga$_{1-x}$As (solid lines are linear fits).}
\label{fig3:AlGaAs}
\end{figure}

It is found that for samples with \emph{x}=0 and 0.1, although $\phi_{V}$ can be readily measured both by BEES and s-BEES processes, $\phi_{C}$ can not be measured due to the overwhelming internal hole current under the forward $V_{C}$ needed for electron injection. For $x$$\geq$0.2, the measured bandgaps of Al$_{x}$Ga$_{1-x}$As $E_{g}(x)\equiv\phi_{C}(x)+\phi_{V}(x)$ (Fig.~3(a)) agree well with the established values~\cite{vurgaftman:jap:89:5815} (within 2\%). In the direct-gap regime, no obvious band bowing effect was observed~\cite{oelgart:sst:2:468}. In the indirect-gap regime, a bandgap corresponding to L valley instead of the lower lying X valley in the CB is observed. This may be explained by electron scattering in the semiconductor. The calculated mean free path (mfp) for the $\Gamma$, L, and X electrons near the corresponding threshold energy in GaAs at 85~K are 150, 30, and 3~nm~\cite{lee:jvsta:15:1351}, respectively. Although scattering rates may be further surpressed at 4.2~K, the contribution of X electrons is still largely attentuated since their mfp is still much shorter than the combined thickness of the GaAs cap and the barrier layers ($\sim$50~nm). Previous BEEM work on direct-gap GaAs/Al$_{x}$Ga$_{1-x}$As ($x$$\leq$0.42) SB HJs found that the collector current is composed only of $\Gamma$ and L electrons~\cite{kozhevnikov:prl:82:3677}, and the contribution from the off-axis L channel is greater due to interfacial scatterings at the m-s interface.


The main results from the quaternary AlGaInP alloy system are shown in Fig.~4. Fig.~4(b) and (c) show the measured $\phi_{V}$ and $\phi_{C}$ of Al-(Al$_{x}$Ga$_{1-x}$)$_{0.51}$In$_{0.49}$P Schottky contacts as well as the VB and CB offsets $\Delta E_{V}(x)$ and $\Delta E_{C}(x)$ derived the same way as above. $\Delta E_{V}(x)$ shows a linear dependence on \emph{x} for 0$<$x$<$0.7 as $\Delta E_{V}(x)=0.293+0.306x$~(eV) ($\pm3$\%) which can be explained by an argument similar to the case for AlGaAs. For $x$$>$0.7, $\Delta E_{V}(x)$ decreases with \emph{x} as $\Delta E_{V}(x)=0.541-0.037x$ ($\pm4$\%). 

\begin{figure}
\includegraphics[width=0.4\textwidth]{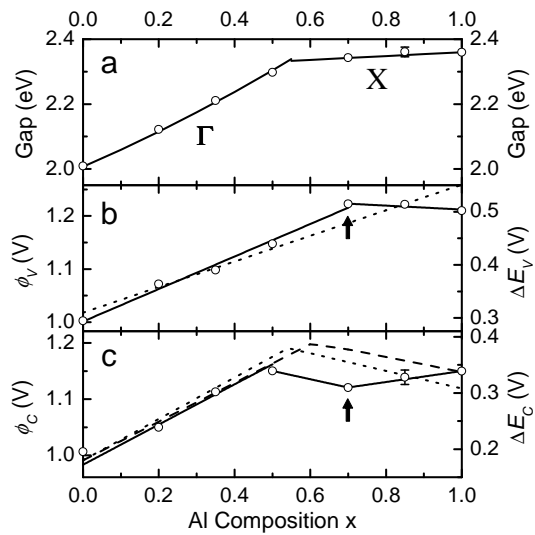}
\caption{Al mole fraction \emph{x} dependence (measured at 4.2 K) of (a) energy bandgap of (Al$_{x}$Ga$_{1-x}$)$_{0.51}$In$_{0.49}$P (solid lines are from Ref.~\cite{vurgaftman:jap:89:5815}), (b) valence-band barrier $\phi_{V}$ and valence-band offset of GaAs/(Al$_{x}$Ga$_{1-x}$)$_{0.51}$In$_{0.49}$P, and (c) conduction-band barrier $\phi_{C}$ and conduction-band offset of GaAs/(Al$_{x}$Ga$_{1-x}$)$_{0.51}$In$_{0.49}$P. Dotted lines and dashed line are deduced from Ref.~\cite{kish:AlGaInPbook} and~\cite{vignaud:jap:93:384}, respectively. Vertical arrows show the transition at \emph{x}$\sim$0.7.}
\label{fig4:AlGaInP}
\end{figure}

The CB offset $\Delta E_{C}(x)$ shows a more complicated behavior. In the direct-gap regime (0$<$$x$$<$0.5), $\Delta E_{C}(x)$ increases linearly with \emph{x} as $\Delta E_{C}=0.172+0.358x$ ($\pm7$\%). The measured $\Gamma$ band offsets agree well with the expected values deduced by the transitivity rule, i.e. adding $\Delta E_{V}=0.31$~eV and $\Delta E_{C}=0.18$~eV of the GaInP-GaAs HJ to the reported $\Delta E_{V}$ and $\Delta E_{C}$ on AlGaInP-GaInP HJs~\cite{kish:AlGaInPbook,vignaud:jap:93:384}, respectively. At $x$$>$0.5, $\Delta E_{C}$ starts to decrease signaling a direct-indirect transition of the CB minima. Surprisingly, $\Delta E_{C}$ increases with \emph{x} again at $x$$>$0.7 as $\Delta E_{C}=0.242+0.097x$ ($\pm1$\%). The measured bandgap values of (Al$_{x}$Ga$_{1-x}$)$_{0.51}$In$_{0.49}$P, after summing $\phi_{V}$ and $\phi_{C}$ (Fig.~4(a)), match well with the established data acquired by optical methods~\cite{vurgaftman:jap:89:5815}(within 0.4\%). The bandgap in the indirect-gap regime is found to be the X valley in the CB. The bandgap does not show a transition at \emph{x}$\sim$0.7 because the opposite trends in the CB and the VB at $x$$>$0.7 are effectively canceled out. The $\Gamma$ band offset ratio \emph{r} for GaAs/AlGaInP HJs is found to increase from 37:63 at x=0 to 44:56 at x=0.5 ($\pm8$\%). Compared with the case for GaAs/AlGaAs, the band offsets at GaAs/AlGaInP HJs with different anions tend to distribute more in the VB, as predicted by the common anion rule~\cite{yu-mccaldin-mcgill:book}.

The nature of the singularity at \emph{x}$\sim$0.7 remains unclear presently and it has not been reported previously. Strain effects should be minimal since all the epilayers were grown at the same temperature and the AlGaInP layers are nearly lattice-matched (with $\leq$0.1\% compressive strain measured by x-ray diffraction) to the GaAs substrates. The ordering effect in AlGaInP alloys can be excluded because it would cause a lowering of the bandgap~\cite{chen:AlGaInPbook}, which is not observed in our case. Intensive studies have been performed on band offsets of AlGaInP-GaInP HJs~\cite{kish:AlGaInPbook}, since it is relevant to the active region of AlGaInP-based orange and yellow light emitters, but little work has been done on the properties of AlGaInP-GaAs interfaces~\cite{watanabe:apl:50:906}.

As a summary, we have demonstrated a self-consistent way to measure the bandgap of a semiconductor and the band offsets in both the conduction band and the valence band of a semiconductor heterojunction. Presently this method is limited to low-temperature measurement to suppress thermally activated currents. A modulation technique may allow measurement at higher temperatures and probing the bandgap of a homogeneous semiconductor without embedded potential steps.

This work was supported by a DARPA HUNT contract No. 222891-01 sub-award from the University of Illinois at Urbana-Champaign, the NSF-funded Nanoscale Science and Engineering Center (NSEC), and the Center for Nanoscale Systems (CNS) at Harvard University.

\end{document}